\documentclass{aa}
\usepackage{graphicx,natbib}
\bibpunct{(}{)}{;}{a}{}{,}

\newcommand{\nh}{N_\mathrm{H}}
\newcommand{\lhx}{L_\mathrm{hx}}
\newcommand{\lx}{L_\mathrm{x}}

\newcommand{\beq}{\begin{equation}}
\newcommand{\eeq}{\end{equation}}
\newcommand{\beqa}{\begin{eqnarray}}
\newcommand{\eeqa}{\end{eqnarray}}

\begin{document}

\title{Discovery of heavily-obscured AGN among 7 INTEGRAL hard X-ray
 sources observed by Chandra} 

\author{S.~Sazonov\inst{1,2} \and M.~Revnivtsev\inst{1,2} \and
 R.~Burenin\inst{2} \and  E.~Churazov\inst{1,2} \and
 R.~Sunyaev\inst{1,2} \and W.R. Forman\inst{3} \and
 S.S. Murray\inst{3}}

\offprints{sazonov@mpa-garching.mpg.de}

\institute{Max-Planck-Institut f\"ur Astrophysik,
           Karl-Schwarzschild-Str. 1, D-85740 Garching bei M\"unchen,
           Germany
     \and   
           Space Research Institute, Russian Academy of Sciences,
           Profsoyuznaya 84/32, 117997 Moscow, Russia
     \and
           Harvard-Smithsonian Center for Astrophysics, 60 Garden St.,
           Cambridge, MA 02138, USA 
}
\date{Received / Accepted}

\authorrunning{Sazonov et al.}
\titlerunning{Identification of hard X-ray sources}

\abstract  
{}
  {We identify hard X-ray sources discovered by the
  INTEGRAL all-sky survey. We complete identification of a unique
  sample of active galactic nuclei (AGN) selected in the hard X-ray
  band (17--60~keV) with minimal effects from
  absorption. Subsequently, we determine the fraction of obscured AGN
  in the local Universe.} 
  {We observed 7 INTEGRAL sources with the Chandra X-ray Observatory to
  refine their localization to $\sim$2~arcsec and to study their
  X-ray spectra.} 
  {Two sources are inferred to have a Galactic origin:
  IGR~J08390$-$4833 is most likely a magnetic cataclysmic variable
  with a white dwarf spin period $\sim$1,450~s; and IGR~J21343+4738 is
  a high-mass X-ray binary. Five sources (IGR~J02466$-$4222,
  IGR~J09522$-$6231, IGR~J14493$-$5534, IGR~J14561$-$3738, and
  IGR~J23523+5844) prove to be AGN with significant intrinsic X-ray
  absorption along the line of sight. Their redshifts and hard X-ray 
  (17--60~keV) luminosities range from 0.025 to 0.25 and from $\sim
  2\times 10^{43}$ to $\sim 2\times 10^{45}$~erg~s$^{-1}$, respectively,
  with the distance to IGR~J14493$-$5534 remaining
  unknown. The sources IGR~J02466$-$4222 and IGR~J14561$-$3738 are likely
  Compton-thick AGN with absorption column densities
  $\nh>10^{24}$~cm$^{-2}$, and the former further appears to be one of
  the nearest X-ray bright, optically-normal galaxies.} 
  {With the newly-identified sources, the number of heavily-obscured ($\nh\ga  
  10^{24}$~cm$^{-2}$) AGN detected by INTEGRAL has increased to
  $\sim$10. Therefore, such objects constitute 10--15\% of hard
  X-ray bright, non-blazar AGN in the local Universe. 
  The small ratio ($\ll 1$\%) of soft (0.5--8.0 keV) to hard (17--60
  keV) band fluxes (Chandra to INTEGRAL) and the non-detection of
  optical narrow-line emission in some of the Compton-thick AGN in our
  sample suggests that there is a new class of objects in which the
  central massive black hole may be surrounded by a
  geometrically-thick dusty torus with a narrow ionization cone.}


\maketitle

\section{Introduction}

Our analysis of four years of all-sky observations by the
IBIS/ISGRI soft gamma-ray imager \citep{ubeetal03} aboard the
INTErnational Gamma-Ray Astrophysics Laboratory (INTEGRAL,
\citealt{winetal03}) has provided a large ($>130$) sample of mostly 
nearby ($z\la 0.1$) active galactic nuclei (AGN) detected in the
17--60~keV energy band \citep{krietal07}. These AGN
populate the whole range of X-ray absorption column densities, 
from unobscured ($\nh<10^{22}$~cm$^{-2}$) to Compton-thick ($\nh\ga
10^{24}$~cm$^{-2}$) sources \citep{sazetal07}. Thus, the
INTEGRAL survey together with the survey by Swift
\citep{ajeetal08,tueetal08} for the first time permit a statistical
investigation of the properties of the local AGN population in a
nearly unbiased manner (except for very Compton-thick sources, $\nh\ga
10^{25}$~cm$^{-2}$), thanks to the detection of sources at
energies above the photoabsorption cutoff in their spectra
($\sim$10~keV).
 
Almost all INTEGRAL sources out of the Galactic plane ($|b|>5^\circ$)
have already been identified \citep{krietal07,biretal07}. This has
allowed us to measure the hard X-ray (17--60~keV) luminosity function
and distribution of absorption column densities for nearby AGN
\citep{sazetal07}. We have also constructed the cumulative spectral
energy distribution of local AGN in the 3--300~keV energy range and,
by comparing it with the spectrum of the cosmic X-ray background,
imposed interesting, new constraints on AGN evolution \citep{sazetal08}.

Nonetheless, several tens of INTEGRAL sources remain unidentified,
most of them located near the Galactic plane ($|b|<5^\circ$) and in
the Galactic Center region, where searches for X-ray and optical
counterparts within the few-arcmin INTEGRAL localization regions are
complicated in comparison with the extragalactic sky. Based on the
INTEGRAL AGN counts at $|b|>5^\circ$ \citep{krietal07}, we 
may expect that a significant fraction of the as yet unidentified sources at
low Galactic latitudes should be AGN, including
heavily-obscured sources ($\nh\ga 10^{24}$~cm$^{-2}$). There is a lot of
interest in searching for such objects, as their census is not yet
complete even in the local Universe, let alone at redshift
$z\ga 0.1$, and we do not yet fully understand their nature. 

We 
previously used the Chandra X-ray Observatory, with its excellent angular
resolution and sensitivity, to refine the positions of 8 INTEGRAL
sources to a few arcseconds and to measure their X-ray spectra. This
allowed us to identify 5 of them as nearby AGN with absorption column
densities up to $\sim 10^{24}$~cm$^{-2}$ \citep{sazetal05}. Here we
present Chandra observations of another 7 INTEGRAL sources and
demonstrate that most of these also are obscured AGN. A cosmology with
$\Omega_{\rm m}=0.3$, $\Omega_\Lambda=0.7$, and
$H_0=75$~km~s$^{-1}$~Mpc$^{-1}$ is adopted throughout the paper. All
quoted uncertainties are 1$\sigma$ unless noted otherwise. 

\section{Observations and data analysis}

For the present study, we compiled a sample of 7 new hard X-ray
sources detected with greater than 5$\sigma$ significance in the
summed INTEGRAL sky map. Six of the sources are included in the
catalog of \cite{krietal07} and one (IGR~J08390$-$4833) was discovered
after the catalog publication. These sources are located
significantly far away from both the Galactic plane ($>2^\circ$) 
and Galactic Center ($>30^\circ$), and were not detected in soft
X-rays by the ROSAT All-Sky Survey \citep{vogetal99}. The IBIS/ISGRI
localization regions are $\sim 3^\prime$ in radius (90\% confidence). This
INTEGRAL sample was observed by the Chandra Advanced CCD Imaging
Spectrometer (ACIS-I) in December 2006--January 2007, with an exposure of
$\sim$3.5~ks per source.  

Due to the anticipated brightness of the INTEGRAL sources -- typical fluxes
of these sources in the 17--60~keV energy band are
$\sim10^{-11}$~erg~s$^{-1}$~cm$^{-2}$, which for unabsorbed spectra
would correspond to $\sim$1~cnt~s$^{-1}$ in the ACIS detectors --  
we undertook special efforts to avoid photon pileup in the ACIS
CCDs. Specifically, the sources were intentionally observed with a
$\sim10^\prime$ offset with respect to the optical axis of the X-ray
telescope, which resulted in a distribution of X-ray counts from the
target source over a wide area ($\sim10\arcsec$ in
radius) and prevented pileup problems. Due to this 
observational scheme, the accuracy of determination of source
positions has decreased to 2--2.5$\arcsec$.

We reduced the data following a standard procedure fully described in
\cite{viketal05}. The detector background was modeled with the stowed dataset
(http://cxc.harvard.edu/contrib/maxim/stowed). We detected point
sources with the wavelet decomposition package wvdecomp of ZHTOOLS 
\citep{viketal98}\footnote{http://hea-www.harvard.edu/saord/zhtools/}.
The spectral modeling was done with XSPEC \citep{arnaud96}.

\section{Results}

\begin{table*}[htb]
\caption{Localization and identification of INTEGRAL sources
\label{tab:ident}
}
\begin{tabular}{lrrclcl}
\hline
\hline
\multicolumn{2}{c}{INTEGRAL} &
\multicolumn{1}{c}{Chandra position} &
\multicolumn{1}{c}{Offset} &
\multicolumn{1}{c}{Identification} &
\multicolumn{1}{c}{Type} &
\multicolumn{1}{c}{$z$}
\\
\cline{1-2}
\multicolumn{1}{c}{Name} &
\multicolumn{1}{c}{$\alpha$, $\delta$ (2000)} &
\multicolumn{1}{c}{$\alpha$, $\delta$ (2000)} &
\multicolumn{1}{c}{(')} &
\multicolumn{1}{c}{} &
\multicolumn{1}{c}{} &
\multicolumn{1}{c}{}
\\
\hline
IGR J02466$-$4222 &  41.644 $-$42.360 & 02 46 36.91 $-$42 21 59.0  &
0.6 & MCG -07-06-018  & AGN & 0.0696\\
IGR J08390$-$4833 & 129.728 $-$48.556 &  08 38 48.98 $-$48 31 25.4 &
2.2 & USNO-B1.0 0414-0125587 & magn. CV? &\\
IGR J09522$-$6231 & 148.025 $-$62.523 &  09 52 20.29 $-$62 32 36.1 &
2.1 & & AGN & 0.252 \\
IGR J14493$-$5534 & 222.311 $-$55.589 & 14 49 12.79 $-$55 36 21.0 &
1.0 & 2MASX J14491283$-$5536194 & AGN &\\
IGR J14561$-$3738 & 224.055 $-$37.632 & 14 56 08.23 $-$37 38 53.8 &
1.4 & ESO 386-G034 & AGN & 0.0246 \\
IGR J21343+4738 & 323.625 +47.614 & 21 34 20.41 +47 38 01.6 &
2.0 & USNO-B1.0 1376-0511904 & HMXB &\\
IGR J23523+5844 & 358.079 +58.745 & 23 52 21.96 +58 45 31.5 & 
0.9 & USNO-B1.0 1487-0398304 & AGN & 0.163\\
\hline
\end{tabular}
\end{table*}

\begin{figure*}
\centering
\includegraphics[bb=0 0 480 700,width=0.75\textwidth]{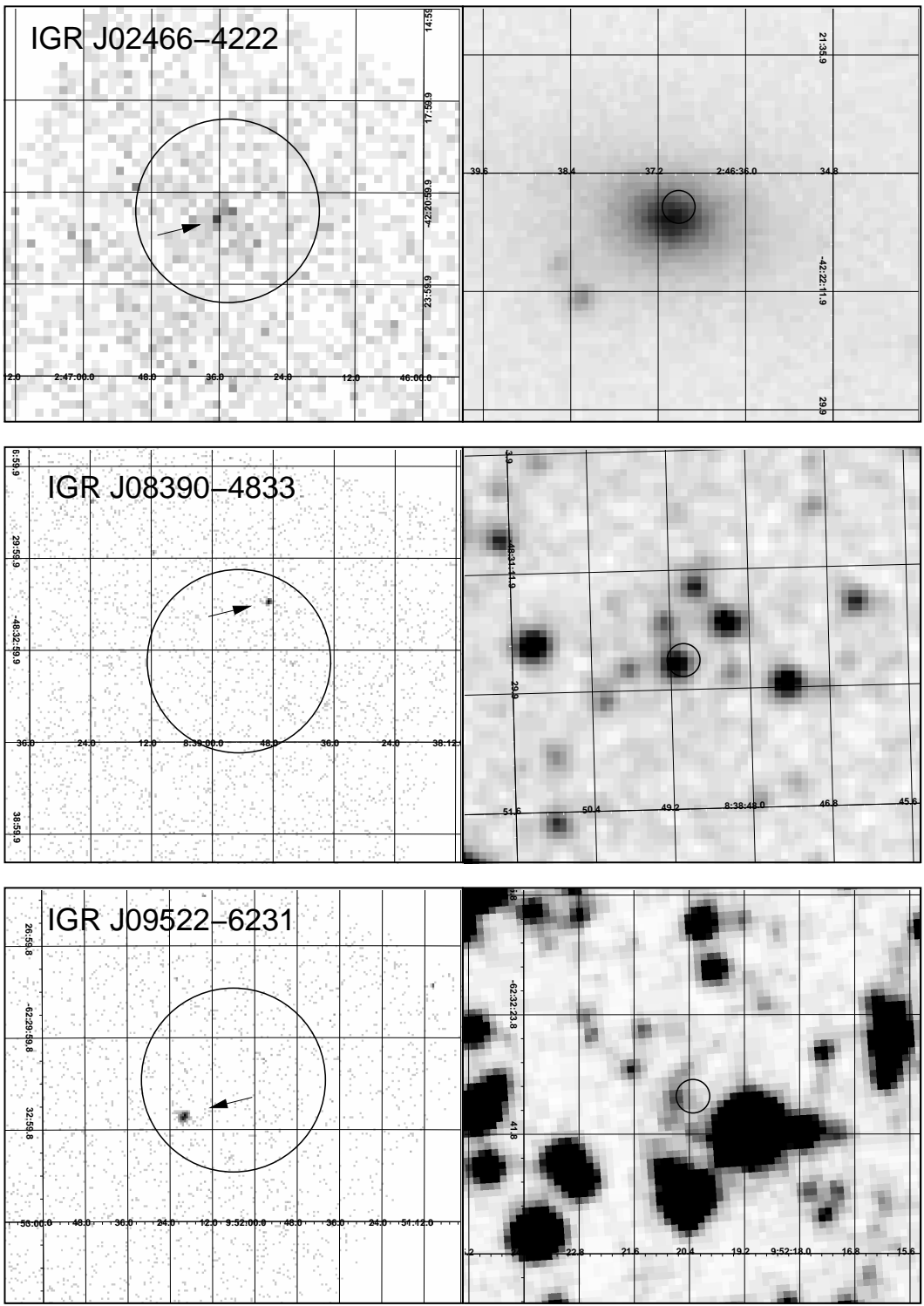}
\caption{{\sl Left panels:} Chandra $\sim$12' images around the INTEGRAL
sources. The 3'-radius circle indicates the estimated INTEGRAL
90\%-confidence localization region. The arrow points to the likely
X-ray counterpart. {\sl Right panels:} Optical (DSS II red) or
near-infrared (2MASS, for IGR~J14493$-$5534 and IGR~J14561$-$3738
only) $\sim$1' images aroung the Chandra counterparts. The 2.5''-radius circle
indicates the estimated uncertainty in the X-ray source position.
}
\label{fig:images}
\end{figure*}

\setcounter{figure}{0}
\begin{figure*}
\centering
\includegraphics[bb=0 0 480 700,width=0.75\textwidth]{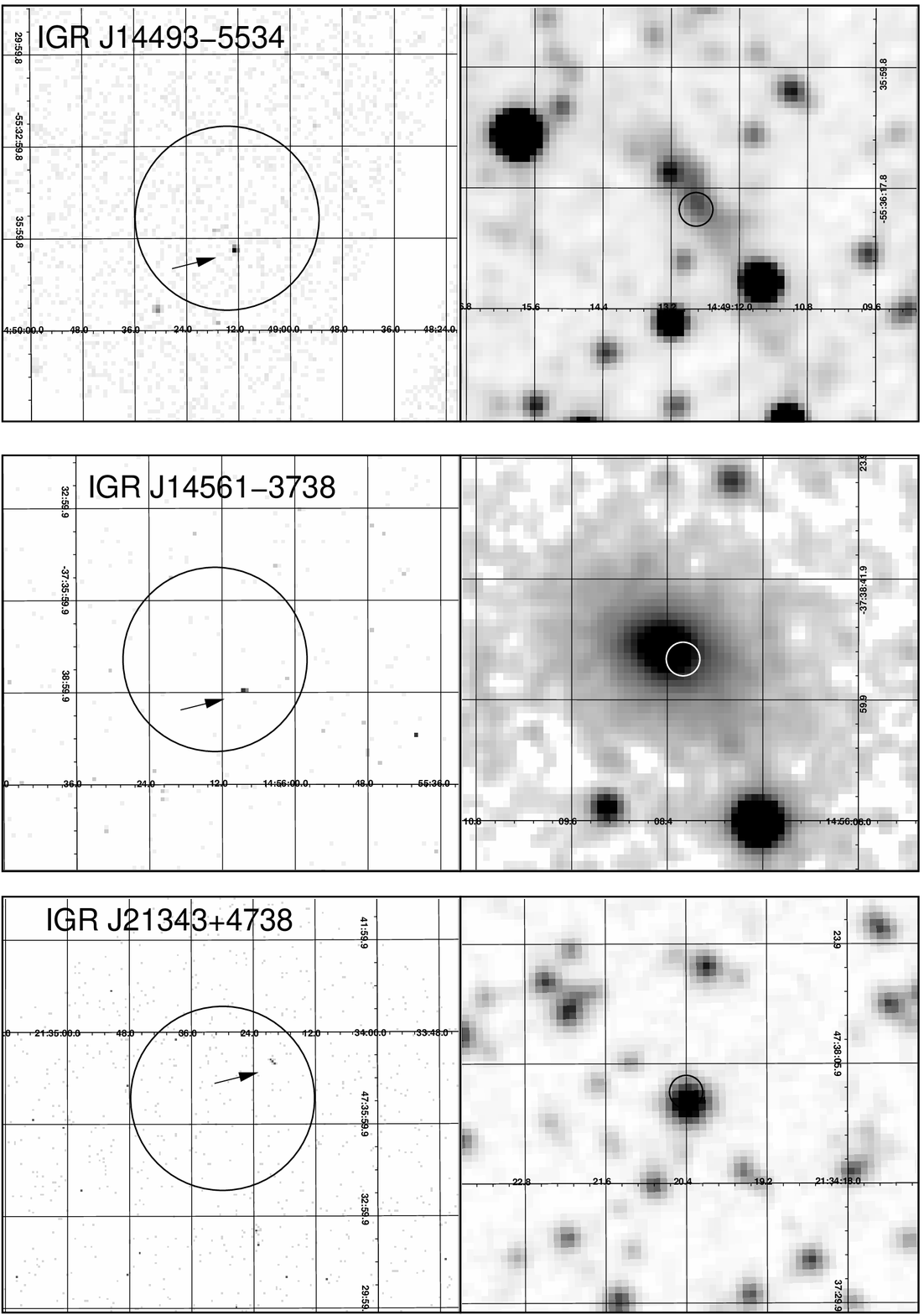}
\caption{Continued.
}
\end{figure*}

\setcounter{figure}{0}
\begin{figure*}
\centering
\includegraphics[bb=0 0 490 230,width=0.75\textwidth]{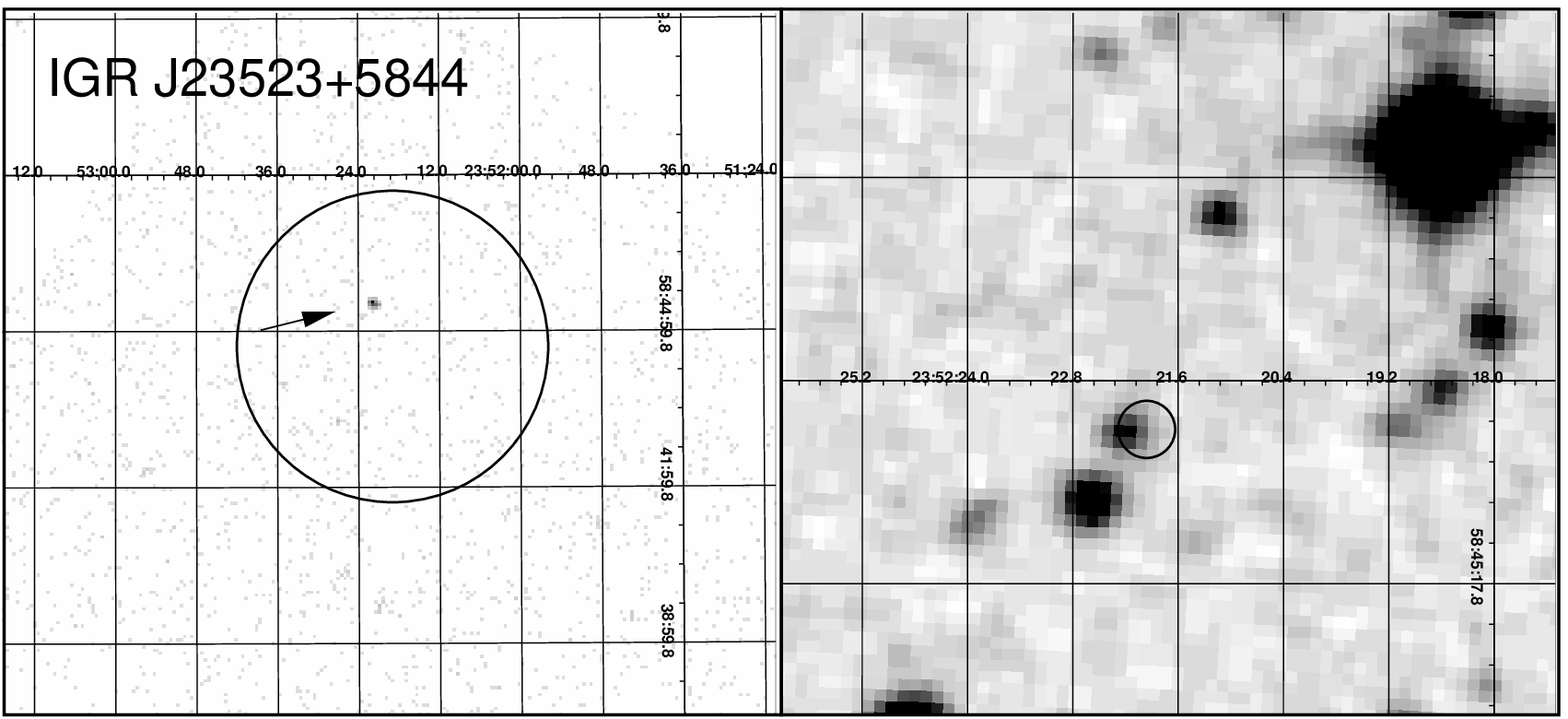}
\caption{Continued.
}
\end{figure*}

\begin{figure*}
\centering
\includegraphics[bb=0 320 590 720,width=\textwidth]{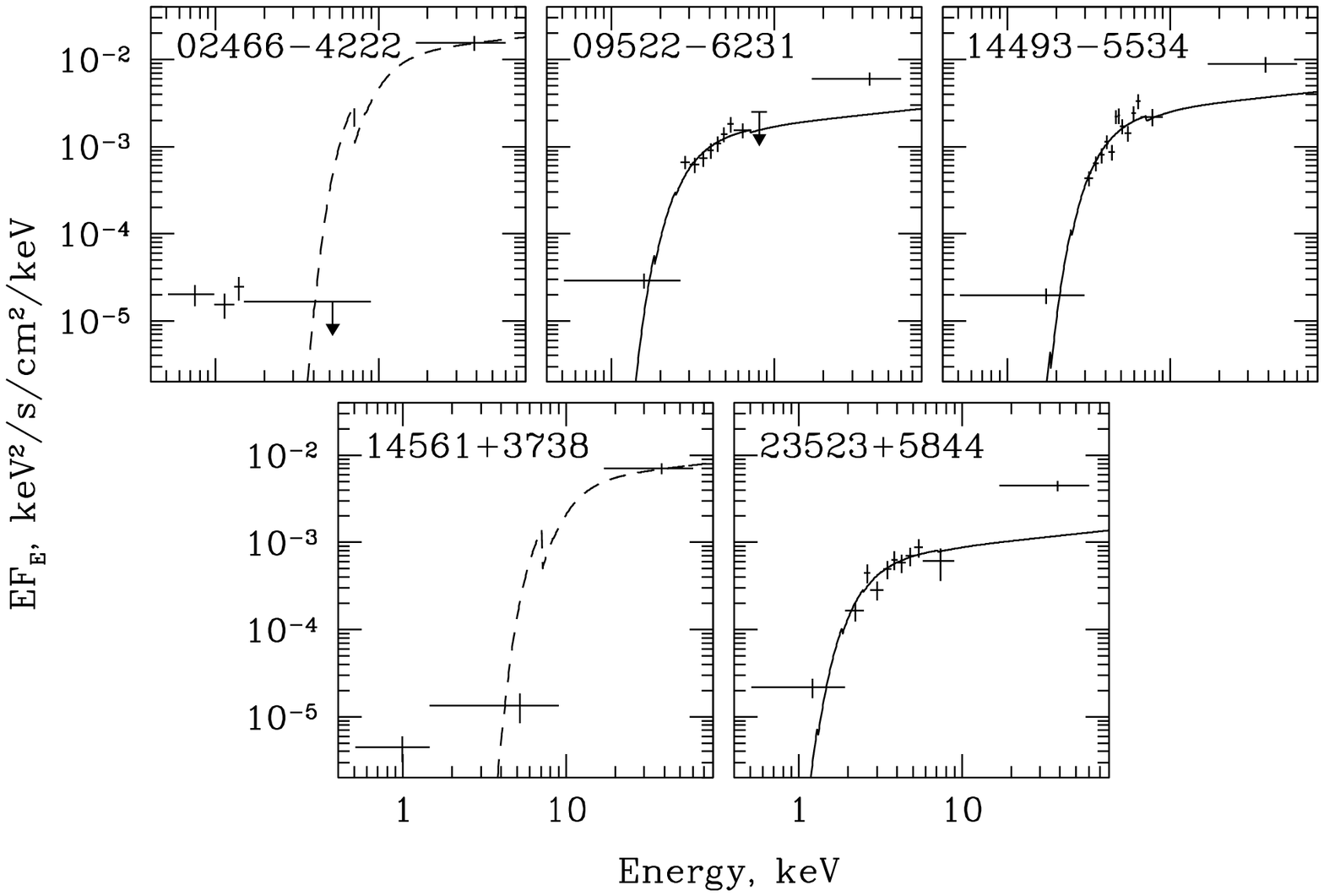}
\caption{X-ray spectra of the identified AGN obtained by Chandra (data
from 0.5--9~keV) and INTEGRAL (data at 17--60~keV). 
IGR source names are given in the left upper corner of the panels. The
solid lines represent the best-fit models to the Chandra data by an
absorbed power-law spectrum with photon index $\Gamma=1.8$ and
absorption column densities as quoted in Table~\ref{tab:spec_agn}. For
IGR~J02466$-$4222 and IGR~J14561+3738, the dashed line represents the
power-law ($\Gamma=1.8$) spectrum absorbed by neutral material with
column density $\nh=10^{24}$~cm$^{-2}$ that matches the INTEGRAL hard
X-ray flux.}
\label{fig:spectra_agn}
\end{figure*}

\begin{figure*}
\centering
\includegraphics[bb=0 400 590 690,width=\textwidth]{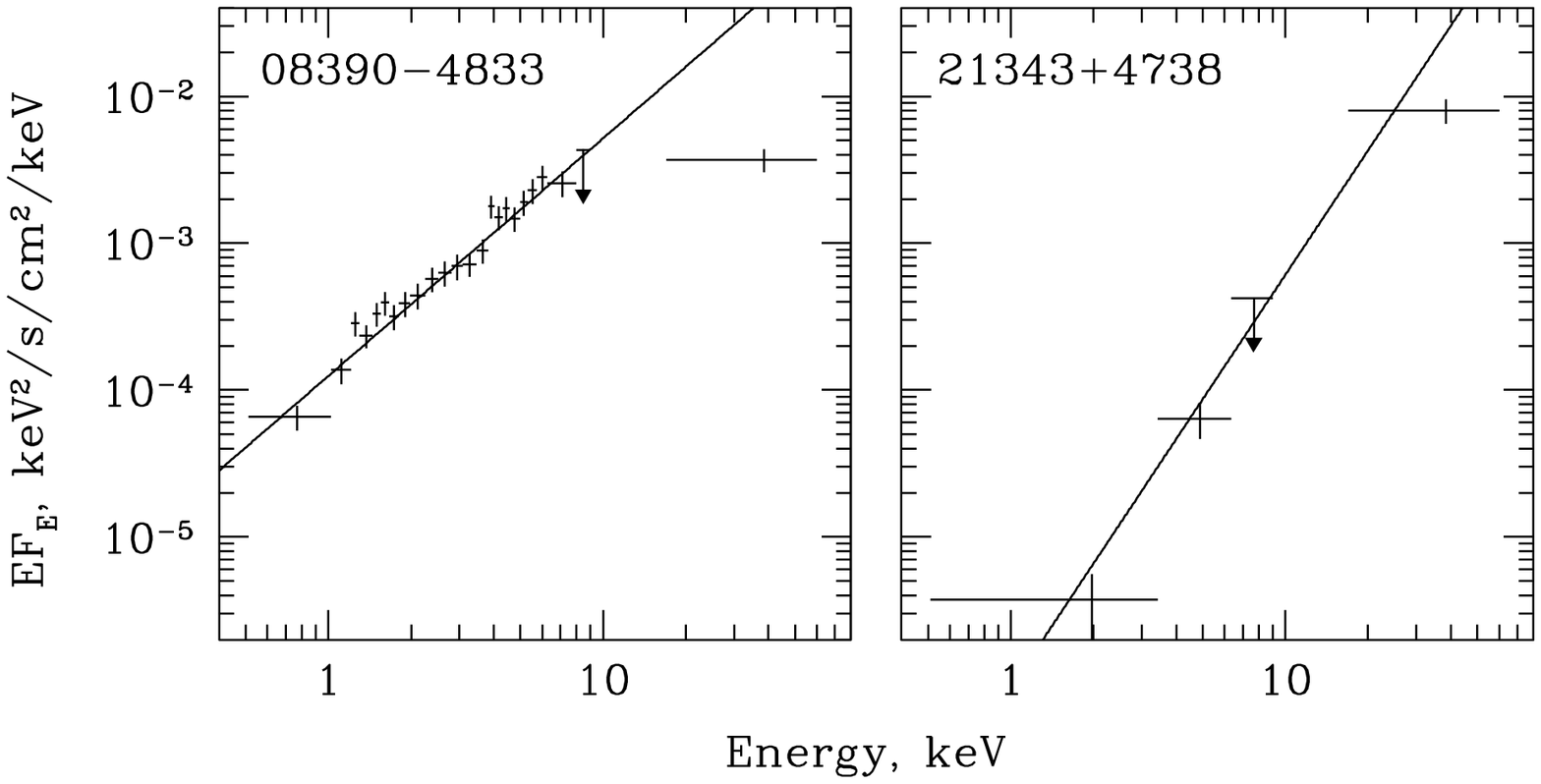}
\caption{X-ray spectra of the identified Galactic sources obtained by
Chandra (data from 0.5--9~keV) and INTEGRAL (data  at 17--60~keV). IGR
source names are given in the left upper corner of the
panels. The solid lines represent the best-fit power- law models of
the Chandra data, with the photon index values given in
Table~\ref{tab:spec_gal}.
}
\label{fig:spectra_gal}
\end{figure*}

In Fig.~\ref{fig:images}, we present the Chandra X-ray images 
around the INTEGRAL sources and optical
(DSS\footnote{http://archive.eso.org/dss/dss/}) or near-infrared
(2MASS\footnote{http://www.ipac.caltech.edu/2mass/}) 
images around their likely X-ray counterparts. The Chandra positions,
their offsets from the IBIS/ISGRI ones and the resulting
identifications of the INTEGRAL sources are listed in
Table~\ref{tab:ident}. In Figs.~\ref{fig:spectra_agn} and 
\ref{fig:spectra_gal}, we show the broadband (0.5--60~keV) X-ray
spectra obtained by Chandra and INTEGRAL for those sources, which we
conclude to be AGN and Galactic objects, respectively. Based on these
spectra, we have estimated or constrained such properties of the
sources as spectral slopes, line-of-sight absorption column densities,
X-ray, and hard X-ray fluxes and luminosities. This information is
collected in Tables~\ref{tab:spec_agn} and \ref{tab:spec_gal} for
the extragalactic and Galactic objects, respectively. Below we
describe our results on a source-by-source basis. 

\subsection{IGR~J02466$-$4222 -- a Compton-thick AGN/X-ray bright,
optically-normal galaxy?}

In the Chandra field (see Fig.~\ref{fig:images}), there is a
faint ($\sim$20 counts) X-ray source inside the IBIS/ISGRI localization
region. The position of the source  
is consistent with the center of galaxy MCG~-07-06-018, located at
redshift $z=0.0696$ (as given in the NASA/IPAC Extragalactic Database,
NED\footnote{http://nedwww.ipac.caltech.edu/}). According to the
HyperLeda database\footnote{http://leda.univ-lyon1.fr/}, this is a
giant ($M_B=-22.64$), early-type (E-S0) galaxy and 
the brightest galaxy in the cluster AS~0296 \citep{gonetal05}.  

Assuming a Crab-like spectrum, the Chandra source has flux
$\sim 7\times 10^{-14}$~erg~s$^{-1}$~cm$^{-2}$ 
in the 0.5--8~keV band. The object is located in the extragalactic sky
($l=253.5^\circ$, $b=-61.1^\circ$) where the expected number density of
X-ray sources with similar or higher fluxes is $\sim$20 per sq.~deg
(see, e.g., \citealt{giletal07}). Thus, there is a significant, $\sim$15\%,
probability of finding by chance a source of such brightness within the
IBIS/ISGRI 90\%-confidence region. However, we should also consider
the fact that giant galaxies like MCG~-07-06-018 are very
rare. According to the galaxy luminosity function measured at $z=0$ 
(e.g., \citealt{noretal02}), there are just $\sim$100 galaxies with 
$M_B<-22.5$ within the local volume of the Universe at 
$z<0.1$. Thus, the probability of finding by chance a giant galaxy like
MCG~-07-06-018, within 3~arcmin of one of the $\sim$400
\citep{krietal07} INTEGRAL sources, is only $\sim$1\%.

Therefore, IGR~J02466$-$4222 is most likely associated
with MCG~-07-06-018. We can then estimate the luminosity of the 
source in the 17--60~keV energy band from the flux measured by IBIS/ISGRI (see
Table~\ref{tab:spec_agn}) at $\lhx\sim 3\times 10^{44}$~erg~s$^{-1}$. For
comparison, the X-ray luminosity of the Chandra X-ray counterpart is
only $\lx\sim 8\times 10^{41}$~erg~s$^{-1}$ (0.5--8~keV). This
suggests that IGR~J02466$-$4222 may be a heavily-obscured,
intrinsically-luminous AGN.

It is impossible to perform a detailed analysis of the X-ray spectrum
based on the $\sim$20 photons detected by Chandra from the
source. However, a comparison of the flux measured by INTEGRAL in the
17--60~keV band with the Chandra counts at energies below 10~keV (see
Fig.~\ref{fig:spectra_agn}) suggests that the material obscuring
direct X-ray emission from IGR~J02466$-$4222 must have column density
$\nh>10^{24}$~cm$^{-2}$, unless the source has faded 
by more than a factor of $\sim$5 
between the INTEGRAL observations in June 2006 and the Chandra
observation in January 2007. 

The weak X-ray emission detected by Chandra is spatially unresolved,
which implies that it originates within 
$\sim$15~kpc of the MCG -07-06-018 nucleus. Also, it clearly
represents a separate, relatively soft spectral component unaffected
by absorption (see Fig.~\ref{fig:spectra_agn}). By analogy with
well-studied Compton-thick AGN such as NGC~1068, NGC~4945, and others
(see, e.g., \citealt{matetal00}, and references therein), and taking into
account that MCG -07-06-018 is a giant galaxy at the center of a
galaxy cluster, the Chandra source may be a combination 
of reflected and/or scattered emission from the active nucleus,
superposition of point X-ray sources in the galaxy, and emission
from hot gas peaked on the central cluster galaxy. Significantly
longer X-ray observations could clarify the origin of the Chandra
X-ray emission. 

In summary, IGR~J02466$-$4222 is most likely a Compton-thick AGN. To
further investigate its nature, we analyzed the optical spectrum of
MCG~-07-06-018 (=6dF~J0246370$-$422201) from the 6dF
survey\footnote{http://www.aao.gov.au/local/www/6df/}. Surprisingly, 
it looks like a typical spectrum of an early-type galaxy with no emission lines
that would indicate the presence of an active nucleus. We can put a 3$\sigma$
upper limit on the luminosity of an [OIII]$\lambda$5007 emission line:
$<6\times 10^{41}$~erg~s$^{-1}$. Thus, the ratio of the
hard X-ray (17--60~keV) to the [OIII]$\lambda$5007 luminosities for
IGR~J02466$-$4222 is at least 500 and is thus larger than in at least
$\sim 90$\% of nearby Seyfert galaxies (\citealt{hecetal05}), 
including those discovered by INTEGRAL \citep{biketal06,buretal08}.

These properties suggest that IGR~J02466$-$4222 may be one of the nearest
X-ray bright, optically-normal galaxies (e.g., \citealt{cometal02}) and
similar to another INTEGRAL source, IGR~J13091+1137. The latter is
associated with the optically-normal galaxy NGC~4992 at $z=0.025$
\citep{masetal06a} and is also a nearly Compton-thick AGN ($\nh\sim
10^{24}$~cm$^{-2}$, \citealt{sazetal05}).  

\subsection{The source IGR~J08390$-$4833}

There is a bright Chandra source inside the IBIS/ISGRI
localization region (see Fig.~\ref{fig:images}). The sky 
density of X-ray sources with similar 2--10~keV fluxes ($\sim
2\times 10^{-12}$~erg~s$^{-1}$~cm$^{-2}$) is $\sim
0.2$~deg$^{-2}$ in the extragalactic sky and typically $\sim
1$~deg$^{-2}$ in the Galactic plane \citep{sugetal01}. Therefore,
since the object is located not far from the Galactic plane
($l=266.6^\circ$, $b=-4.3^\circ$), the probability of a chance
positional coincidence of the Chandra and INTEGRAL sources is $\sim
10^{-3}$--$10^{-2}$. Therefore, the Chandra source is most
likely the X-ray counterpart of IGR~J08390$-$4833.

The spectrum measured in the 0.5--9~keV energy band (see
Fig.~\ref{fig:spectra_gal}) can be well fit ($\chi^2=558.3$ for 579
d.o.f.) by a power law with photon index $\Gamma=0.38\pm 0.07$. A
model of a power-law spectrum absorbed by 
neutral gas (with column density $\nh$) provides a somewhat better
fit, with $\Gamma=0.54\pm 0.13$ and $\nh=(1.2\pm 0.9)\times
10^{21}$~cm$^{-2}$. We can put a 3$\sigma$ upper limit on the $\nh$
value of $4\times 10^{21}$~cm$^{-2}$. This is less than the
absorption column density through the whole Galaxy in 
the direction of the source: $N_{\rm H,Gal}\sim 7\times 10^{21}$ and
$\sim 5\times 10^{21}$~cm$^{-2}$, according to \cite{dicloc90} and
\cite{kaletal05}, respectively. This hints at a Galactic origin of
IGR~J08390$-$4833.

A comparison of the Chandra spectrum with the hard X-ray flux measured
by INTEGRAL (see Fig.~\ref{fig:spectra_gal}) indicates that the spectrum
must steepen between $\sim 8$~keV and $\sim 20$--60~keV,
unless the source experienced a strong outburst during the Chandra
observation. 

\begin{figure}
\centering
\includegraphics[width=\columnwidth]{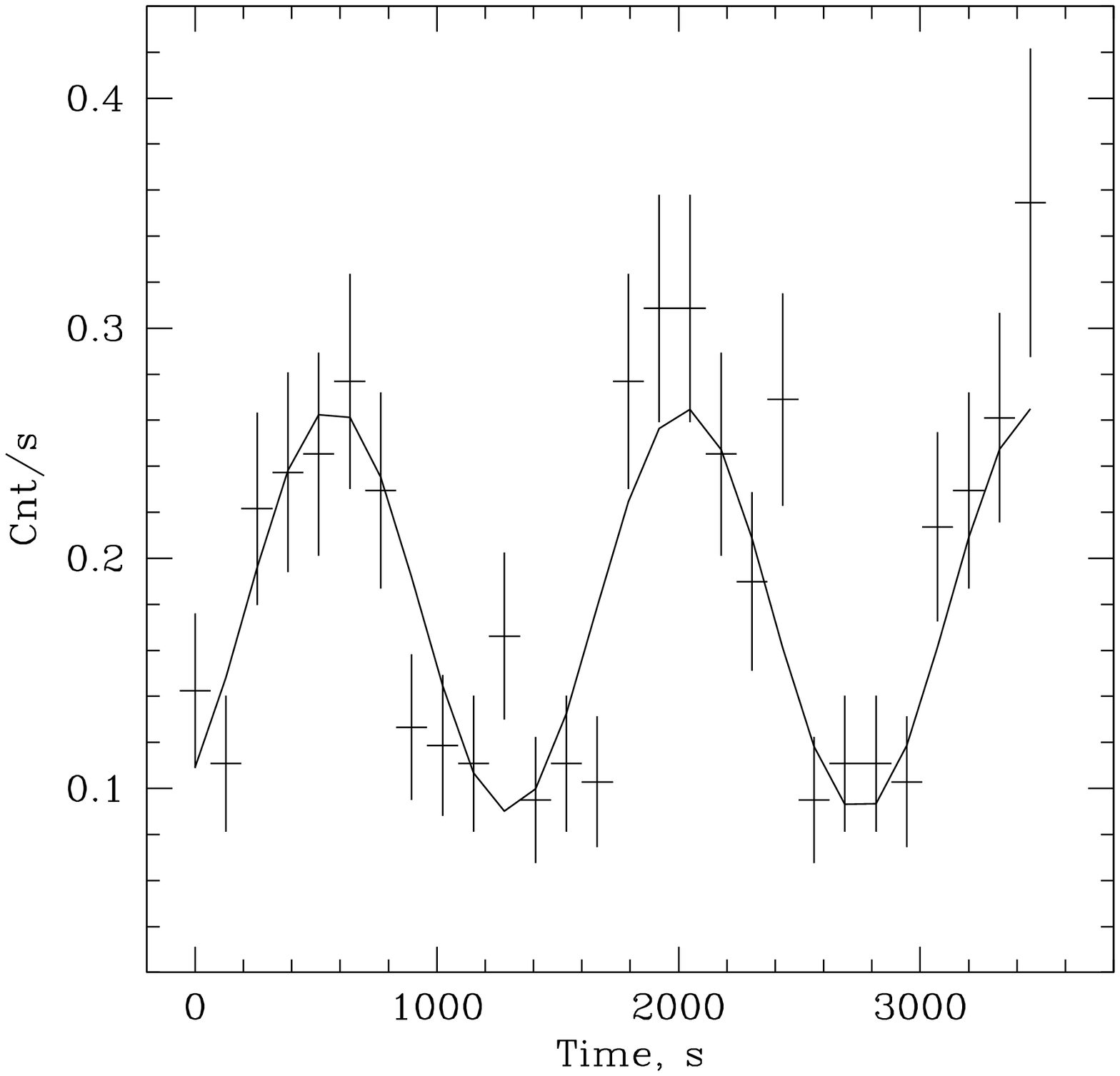}
\caption{Count rate history of IGR~J08390$-$4833 recorded by Chandra,
binned into 28 intervals of 128~s duration. The solid line is the
best fitting model by a sine wave with a period of 1,450~s plus a
constant.
}
\label{fig:igr08390_lc}
\end{figure}

The Chandra count rate history of IGR~J08390$-$4833 
(Fig.~\ref{fig:igr08390_lc}) indicates that the source may be
pulsating. The light curve can be fit fairly well ($\chi^2=34.5$ for
24 d.o.f.) by a model consisting of a constant and a sine wave with
period $P=1450\pm 40$~s. For comparison, fitting to a constant yields
$\chi^2=106.3$ for 27 d.o.f. The inferred pulse
fraction $(C_{\rm max}-C_{\rm min})/(C_{\rm max}+C_{\rm min})\approx
50$\%, where $C_{\rm max}$ and $C_{\rm min}$ are the maximum and
minimum intensities, respectively. No other periodicities have
been found by an epoch folding analysis.

Optical and near-infrared catalogs contain a star, USNO-B1.0
0414-0125587, whose position is consistent with that of the Chandra
source (see Fig.~\ref{fig:images}). According to the USNO-B1 and 2MASS
catalogs, this likely optical counterpart of IGR~J08390$-$4833 is
characterized by the following apparent magnitudes: $B=17.65$,
$R=16.05$, $I=16.83$, $J=15.68$, $H=15.29$, and $K=14.52$.

The very hard ($\Gamma\sim 0.5$) X-ray spectrum at energies below $\sim$10~keV,
with a high-energy cutoff at $\sim 20$~keV, and the large smooth
oscillations of X-ray emission, strongly suggest that 
IGR~J08390$-$4833 is either a high-mass X-ray binary (HMXB)/X-ray
pulsar or a magnetic cataclysmic variable (CV) with the compact
object (a neutron star or a white dwarf, respectively) rotating with a
period $\sim$1,450~s. We, therefore, now discuss these two possibilities.  

In the HXMB scenario, given the weakness of the optical counterpart
($R\sim 16$) and taking into account that the interstellar extinction
toward the source is at most $A_V\approx 2.6$ \citep{schetal98},
IGR~J08390$-$4833 must be located at least $\sim 10$~kpc from us if
this is a $B[e]$/neutron-star binary and much further away if it is a
supergiant X-ray binary (see \citealt{negueruela04} for a review of
the different populations of HMXBs). In this case, the absence of
significant absorption in the X-ray spectrum of IGR~J08390$-$4833
would be only marginally consistent with the substantial HI column
density through the Galaxy in that direction. Furthemore, at 
$(l,b)=266.6^\circ, -4.3^\circ$ the object would be located $>800$~pc
above the Galactic plane, i.e. extraordinarily high for a Galactic
HMXB \citep{grietal02}. We thus find the HMXB scenario unlikely. 

All the available data seem to be more consistent with
IGR~J08390$-$4833 being a magnetic CV. First, the suggested period 
$\sim$~1,450~s is typical of intermediate polar spin periods, a
subclass of magnetic CVs in which the white dwarf 
rotates rapidly compared to the orbital motion (e.g., \citealt{patterson94}). 
Secondly, the broadband X-ray spectrum of IGR~J08390$-$4833 measured
by Chandra and INTEGRAL is similar to those of well-studied magnetic
CVs, which are interpreted in terms of multitemperature optically-thin
thermal emission with $kT\la 30$~keV produced in an accretion column
at the white dwarf surface, partially absorbed in the
surrounding accretion flow (e.g., \citealt{suletal05}). We note though
that we do not detect any iron $K_\alpha$ emission lines at 6.4--7.0~keV
in the Chandra spectrum of IGR~J08390$-$4833, such as usually
observed in the spectra of magnetic CVs. However, the inferred upper
limit on the equivalent width of such a line (obtained by fixing the
line energy and full width at half maximum at 6.7~keV and 600~eV,
respectively), $W_{\rm eq}<300$~eV ($2\sigma$), is consistent with
some of the $W_{\rm eq}$ values measured for polars and intermediate
polars \citep{ezuish99}.

Furthermore, the near-infrared-to-optical spectral energy distribution 
of IGR~J08390$-$4833 constructed from the $BRIJHK$ photometric data
quoted above resembles the similarly constructed spectra
of several magnetic CVs discovered (or co-discovered) by INTEGRAL,
including XSS~J12270$-$4859, IGR~J14536$-$5522, IGR~J15094$-$6649,
IGR~J16167$-$4957, and IGR~J17195$-$4100, which are all $R\sim$14--16 objects
\citep{masetal06b} and, like IGR~J08390$-$4833, are relatively weak
($\sim$mCrab) hard X-ray sources. 

We, therefore, conclude that
IGR~J08390$-$4833 is most likely a magnetic 
CV with a spin period $\sim$1,450~s. Optical spectroscopic
observations could verify this hypothesis by revealing Balmer emission lines 
associated with an accretion disk around a white dwarf. 

\begin{table*}
\begin{center}
\caption{X-ray fuxes, luminosities and absorption column densities of the 
AGN$^{\rm a}$
\label{tab:spec_agn}
}
\begin{tabular}{cccccccc}
\hline
\hline
\multicolumn{1}{c}{Source} &
\multicolumn{1}{c}{$F$ (17--60 keV)} &
\multicolumn{1}{c}{$L$ (17--60 keV)} &
\multicolumn{1}{c}{$F$ (0.5--8 keV)} &
\multicolumn{1}{c}{$L$ (0.5--8 keV)} &
\multicolumn{1}{c}{$N_{\rm H}$} &
\multicolumn{1}{c}{$N_\mathrm{H, Gal}$$^{\rm b}$}
\\
\multicolumn{1}{c}{} &
\multicolumn{1}{c}{$10^{-12}$~erg/s/cm$^{2}$} &
\multicolumn{1}{c}{erg/s} &
\multicolumn{1}{c}{$10^{-12}$~erg/s/cm$^{2}$} &
\multicolumn{1}{c}{erg/s} &
\multicolumn{2}{c}{$10^{22}$~cm$^{-2}$}
\\
\hline
IGR J02466$-$4222 & $31\pm6$ & $(3.2\pm 0.6)\times 10^{44}$ & $\sim0.07$
& $\sim 8\times 10^{41}$ & $>100$ & 0.03\\
IGR J09522$-$6231 & $12\pm 2$ & $(2.0\pm0.4)\times 10^{45}$ & $2.1\pm
0.2$ & ($3.6\pm 0.3)\times 10^{44}$ & $6.3\pm 0.6$ & 0.27\\
IGR J14493$-$5534 & $16\pm2$ & & $2.3\pm 0.2$ &  & $12.3\pm 1.1$ &  0.50\\
IGR J14561$-$3738 & $14\pm3$ & $(1.7\pm 0.3)\times 10^{43}$ &
$\sim0.02$ & $\sim 2\times 10^{40}$ & $>100$ & 0.06\\
IGR J23523+5844 & $8.9\pm1.4$ & $(5.8\pm0.9)\times 10^{44}$ & $1.2\pm 0.1$ &
$(8.4\pm 0.6)\times 10^{43}$ & $3.7\pm 0.5$ & 0.58\\  
\hline
\end{tabular}
\end{center}
$^{\rm a}$ The 17--60~keV fluxes are adopted from the INTEGRAL catalog
\citep{krietal07}. The absorption column densities $\nh$ and 0.5--8~keV
fluxes were determined by fitting the Chandra spectra in the
0.5--9~keV energy band by an absorbed power-law
model with $\Gamma=1.8$, except for IGR~J02466$-$4222 and
IGR~J14561$-$3738, for which only crude estimates of the X-ray fluxes
and lower limits on the $\nh$ values are provided (see main text for 
further explanations). 

$^{\rm b}$ Galactic absorption column from \cite{dicloc90}. 

\end{table*}

\begin{table*}
\begin{center}
\caption{X-ray fluxes and spectral parameters of the Galactic sources$^{\rm a}$
\label{tab:spec_gal}
}
\begin{tabular}{cccccc}
\hline
\hline
\multicolumn{1}{c}{Source} &
\multicolumn{1}{c}{$F$ (17--60 keV)} &
\multicolumn{1}{c}{$F$ (0.5--8 keV)} &
\multicolumn{1}{c}{$\Gamma$} &
\multicolumn{1}{c}{$N_{\rm H}$} &
\multicolumn{1}{c}{$N_\mathrm{H, Gal}$$^{\rm b}$} 
\\
\multicolumn{1}{c}{} &
\multicolumn{1}{c}{$10^{-12}$~erg/s/cm$^{2}$} &
\multicolumn{1}{c}{$10^{-12}$~erg/s/cm$^{2}$} &
\multicolumn{1}{c}{} &
\multicolumn{2}{c}{$10^{22}$~cm$^{-2}$}
\\
\hline
IGR J08390$-$4833 & $7.6\pm1.4$ & $3.5\pm0.3$ & $0.38\pm0.07$ & $<0.4$
($3\sigma$) & 0.71 \\ 
IGR J21343+4738 & $16\pm 3$ & $\sim 0.2$  & $-0.8\pm0.5$ & & 0.32\\ 
\hline
\end{tabular}
\end{center}
$^{\rm a}$ The 17--60~keV flux is adopted from the
INTEGRAL catalog \citep{krietal07} for IGR~J21343+4738 and determined
from INTEGRAL data for IGR~J08390$-$4833. The quoted values of the
power-law index $\Gamma$ and of the 0.5--8~keV flux were obtained by
fitting the Chandra spectra in the 0.5--9~keV energy band by a
power-law model. The upper limit on the $\nh$ value for
IGR~J08390$-$4833 was determined by fitting the Chandra data by an
absorbed power-law model. 

$^{\rm b}$ Galactic absorption column from \cite{dicloc90}. 

\end{table*}

\subsection{The source IGR~J09522$-$6231}

Chandra sees a bright X-ray source inside the IBIS/ISGRI
localization region (see Fig.~\ref{fig:images}).
Its X-ray spectrum (see Fig.~\ref{fig:spectra_agn}) can
be well fit ($\chi^2=434.7$ for 579 d.o.f.) by an 
absorbed power-law model with column density $\nh\sim 6\times
10^{22}$~cm$^{-2}$ (for the photon index fixed at
$\Gamma=1.8$), which is much higher than the Galactic absorption
column density in the direction of the source (see 
Table~\ref{tab:spec_agn}). An extrapolation of this spectral model to
higher energies underpredicts the hard X-ray (17--60~keV) flux
measured by IBIS/ISGRI by a factor of 2--3. This difference may result
from the tendency of INTEGRAL (as of any other mission
conducting a  sensitivity limited survey) to detect weak sources when they are
bright as compared to their long-term average flux level. It may also indicate
that the intrinsic X-ray continuum is harder than the assumed
$\Gamma=1.8$ power law, e.g., a Compton reflection component might be
present. We note that the same explanation may also be applicable to
the spectra of two other sources, IGR~J14493$-$5534 and 
IGR~J23523+5844, considered below. 

In the DSS and
2MASS images (see
Fig.~\ref{fig:images}), there is an object with the centroid at ${\rm
RA}, {\rm Dec}=$~09:52:20.42, $-$62:32:35.4 
(J2000), which is consistent with the Chandra position. With $R\sim
20$, the object is barely detected on the DSS~II red plate. Its
near-infrared magnitudes estimated from the 2MASS images are
$J=16.7$, $H=15.4$, $K=14.8$. \citet{masetal08} have observed
this likely optical counterpart with the 3.6 m telescope at the ESO-La
Silla Observatory and identified it as a Seyfert 1.9 galaxy at
$z=0.252$. This implies that the source has a hard X-ray luminosity of
$\sim 2\times 10^{45}$~erg~s$^{-1}$ (see Table~\ref{tab:spec_agn}). This,
together with the significantly absorbed X-ray spectrum, indicates that
IGR~J09522$-$6231 may be considered a type 2 quasar.

\subsection{IGR~J14493$-$5534 -- an obscured AGN}

There is a bright Chandra source inside the IBIS/ISGRI
localization region (see Fig.~\ref{fig:images}). Similar to the case of
IGR~J08390$-$4833, we estimate the probability of a chance positional
coincidence at $\sim 10^{-3}$--$10^{-2}$, depending on
whether the extragalactic or Galactic log~$N$--log~$S$ function is
used (the object is located at $l=319.1^\circ$, $b=3.5^\circ$). This
indicates that the Chandra source is likely the X-ray counterpart of
IGR~J14493$-$5534.

The Chandra source positionally coincides with galaxy
2MASX~J14491283$-$5536194, whose redshift is unknown. The X-ray  
spectrum measured by Chandra (Fig.~\ref{fig:spectra_agn}) can be well
fit ($\chi^2=435.9$ for 579 d.o.f.) by an absorbed power-law model
with column density $\nh\sim 1.2\times 10^{23}$~cm$^{-2}$, which is much higher
than the Galactic absorption column density in the direction of the
source (see Table~\ref{tab:spec_agn}). We conclude that
IGR~J14493$-$5534 is an obscured AGN. We note that IGR~J14493$-$5534
has also been recently observed by the Swift/XRT telescope and the
X-ray spectrum \citep{maletal07} is in good agreement with that
measured by Chandra.

Spectroscopic optical observations are needed to refine the classification  
and determine the redshift of
IGR~J14493$-$5534=2MASX~J14491283$-$5536194. This should be a feasible
task despite the relatively large Galactic extinction in its
direction, $A_V\approx 2.5$ \citep{schetal98}, since the galaxy is 
relatively bright in the near-infrared band ($K=10.7$, 2MASS).    

\subsection{IGR J14561$-$3738 -- a Compton thick AGN}

This source was first detected and localized to within $\sim 1$~deg
during the RXTE 3--20~keV Slew Survey \citep{revetal04} and received
the name XSS~J14562$-$3735. In the $\sim 3^\prime$ IBIS/ISGRI
localization region there is a conspicuous object -- the
spiral (SBa) galaxy ESO~386-G034 at redshift $z=0.0246$
(NED). \citet{masetal08} recently obtained its optical spectrum and found 
a number of emission lines indicating that this is a Seyfert 2 galaxy. 

With Chandra we found a single X-ray source inside the IBIS/ISGRI
error box of IGR~J14561$-$3738, which is located at the center of
ESO~386-G034 (see Fig.~\ref{fig:images}). This confirms that this
Seyfert 2 galaxy is indeed associated with the INTEGRAL source.  

The Chandra source is very weak. The detected $\sim$12 counts correspond
to a flux of $\sim 2\times 10^{-14}$~erg~s$^{-1}$~cm$^{-2}$ in the 
0.5--8~keV band, assuming a Crab-like spectrum. Combining the Chandra
counts with the flux measured by IBIS/ISGRI in the 17--60~keV band
(Fig.~\ref{fig:spectra_agn}), we infer that direct X-rays from the active
nucleus in IGR~J14561$-$3738 are obscured by material with a column
density of at least $10^{24}$~cm$^{-2}$. We therefore conclude that
IGR~J14561$-$3738 is likely a Compton-thick AGN. 

\subsection{IGR~J21343+4738 -- a high-mass X-ray binary}

In the IBIS/ISGRI localization region, there is a fairly weak
($\sim$24~counts) Chandra source, whose position is consistent with
that of a $R\sim 14$ star, USNO-B1.0~1376-0511904 (see
Fig.~\ref{fig:images}). This likely counterpart of IGR~J21343+4738 has
been observed by the Russian-Turkish 1.5-m Telescope (RTT-150) and
identified as a B3 star, suggesting that the system is a high-mass
X-ray binary (HMXB, \citealt{biketal08}). 

As should be expected for a HMXB, the X-ray spectrum of
IGR~J21343+4738 measured by Chandra is very hard
(Fig.~\ref{fig:spectra_gal}). For example, it can be well fit
($\chi^2=558.1$ for 579 d.o.f.) by a power-law
model with $\Gamma=-0.8\pm 0.5$, although the statistics are
insufficient for detailed spectral analysis. We note that the source
was clearly detected by INTEGRAL in several observations from
2002--2004 but not in later observations, suggesting strong 
variability. For this reason, the average INTEGRAL 17--60~keV flux
value presented in Fig.~\ref{fig:spectra_gal} should not be directly
compared with the Chandra spectrum. We refer the reader to the paper
by \citet{biketal08} for a further 
discussion of the identification and nature of IGR~J21343+4738.

\subsection{IGR~J23523+5844 -- an obscured AGN}

There is a bright Chandra source inside the IBIS/ISGRI
localization region (see Fig.~\ref{fig:images}). The position of this
likely X-ray counterpart 
is consistent with that of a faint point-like optical object. It has
been observed by the RTT-150 telescope and identified as a Seyfert 2
galaxy at redshift $z=0.163$ \citep{biketal08}. \citet{masetal08}
independently obtained similar results.

The X-ray spectrum measured by Chandra (Fig.~\ref{fig:spectra_agn}) can be
well fit ($\chi^2=398.4$ for 579 d.o.f.) by an absorbed power-law
model with column density $\nh\sim 4\times 10^{22}$~cm$^{-2}$, which
is much higher than the Galactic absorption column density in the
source direction (see Table~\ref{tab:spec_agn}). We therefore
conclude that IGR~J23523+5844 is a moderately obscured, luminous
Seyfert galaxy. 

We note that  \citet{rodetal08} have recently observed IGR~J23523+5844
with the Swift/XRT telescope. They reported a position of the weak X-ray
counterpart that differs by just $1''$ from the Chandra position. Also the
absorption column density estimated from the Swift/XRT 
spectrum is consistent with the value found in our analysis.
 
\section{Discussion and conclusions}

In line with our expectations, most of the seven INTEGRAL hard X-ray sources
chosen for short follow-up observations with Chandra proved to be
obscured AGN at redshifts from 0.025 to 0.25. These include
IGR~J02466$-$4222, IGR~J09522$-$6231, IGR~J14493$-$5534,
IGR~J14561$-$3738, and IGR~J23523+5844. Of these, IGR~J09522$-$6231,
IGR~J14561$-$3738, and IGR~J23523+5844 have been optically classified as
Seyfert 1.9--2 galaxies. Based on the X-ray data and 6dF optical spectrum,
IGR~J02466$-$4222 appears to be an X-ray bright, optically-normal
galaxy. At redshift $z=0.0696$, it is thus one the nearest known
objects of this enigmatic class. For IGR~J14493$-$5534 the
distance and optical class are not known yet.

These newly-discovered AGN demonstrate two key features of the
INTEGRAL survey: finding AGN in the poorly studied Galactic zone of
avoidance (i.e., at $|b|\la 5^\circ$) and finding heavily-obscured AGN.  
The inferred line-of-sight, X-ray absorption column densities range from
several $10^{22}$~cm$^{-2}$ for IGR~J09522$-$6231 and IGR~J23523+5844,
through $\sim 10^{23}$~cm$^{-2}$ for IGR~J14493$-$5534 to
$>10^{24}$~cm$^{-2}$ for IGR~J02466$-$4222 and 
IGR~J14561$-$3738. Therefore, the latter two objects are most likely  
Compton-thick AGN. These newly-identified sources increase the sample 
of heavily-obscured ($\nh\ga 10^{24}$~cm$^{-2}$) AGN detected by INTEGRAL
to about 10 objects (see \citealt{sazetal07}), with only half of them
having been known before the survey. 

With these and previous discoveries by INTEGRAL and Swift, the census
of nearby heavily-obscured AGN with hard X-ray fluxes higher than 1--2~mCrab
may be regarded as nearly complete for most of the sky, 
except for the Galactic plane and Galactic Center regions, where
additional follow-up observations are needed. 

Therefore, there is now a well-defined, hard X-ray selected sample of
nearby heavily-obscured ($\nh\ga 10^{24}$~cm$^{-2}$) AGN, which
allows us to place a firm lower limit of $\sim 10$--15\% on the
relative fraction of such AGN in the local Universe. The true fraction may be
higher than this estimate due to bias against finding very
Compton-thick AGN ($\nh\ga 10^{25}$~cm$^{-2}$) even at hard X-ray 
energies. The list of known nearby heavily-obscured AGN will continue
growing in the coming years as the INTEGRAL and Swift surveys deepen
their sky coverage.
 
The very low X-ray (0.5--8~keV) to hard X-ray (17--60~keV) flux ratios
determined from Chandra and INTEGRAL observations for the
Compton-thick AGN IGR~J02466$-$4222 and IGR~J14561$-$3738 indicate
that, in both cases, much less than 1\% of the emission from the
nucleus is scattered by the ambient medium. Such objects were
virtually unknown before the INTEGRAL/Swift era. It is possible that
here we are dealing with AGN buried in a very geometrically  
thick torus that obscures most of the sky as observed from the central 
massive black hole \citep{uedetal07}. This could also explain the weakness (or
absence) of optical narrow-line emission in IGR~J02466$-$4222, as
there would be little radiation escaping from the nucleus to ionize
surrounding interstellar medium.

\smallskip
\noindent {\sl Acknowledgments} This research made 
use of the High Energy Astrophysics Science Archive Research Center
Online Service (provided by the NASA/Goddard Space Flight Center), of
the NASA/IPAC Extragalactic Database (operated by the Jet Propulsion
Laboratory, California Institute of Technology), of the VizieR
catalog access tool (CDS, Strasbourg), of the HyperLeda catalog
(operated at the Observatoire de Lyon), and of data products of the Two
Micron All Sky Survey (joint project of the University of 
Massachusetts and the IPAC, funded by the NASA and NSF). The research was
partially supported by DFG-Schwerpunktprogramme SPP 1177, program of the
Russian Academy of Sciences "Origin and Evolution of Stars and
Galaxies" and grant of the President of Russia NSh-1100.2006.2.



\begin{thebibliography}{}

\bibitem[\protect\citeauthoryear{Ajello et al.}{2008}]{ajeetal08}
Ajello, M., Rau, A., Greiner, L., et al. 2008, ApJ, 673, 96 

\bibitem[\protect\citeauthoryear{Arnaud}{1996}]{arnaud96} Arnaud,
K.A. 1996, in Astronomical Data Analysis Software and Systems V,
ed. Jacoby, G., \& Barnes, J., ASP Conf. Series, 101, 17
	
\bibitem[\protect\citeauthoryear{Bikmaev et al.}{2006}]{biketal06}
Bikmaev, I.F., Sunyaev, R.A., Revnivtsev, M.G., \& Burenin, R.A.
2006, Astron. Lett., 32, 221

\bibitem[\protect\citeauthoryear{Bikmaev et al.}{2008}]{biketal08}
Bikmaev, I., et al. 2008, Astron. Lett., in press

\bibitem[\protect\citeauthoryear{Burenin et al.}{2008}]{buretal08}
Burenin, R.A., et al. 2008, Astron. Lett., 34, 403 

\bibitem[\protect\citeauthoryear{Bird et al.}{2007}]{biretal07}
Bird, A.J., Malizia, A., Bazzano, A., et al. 2007, ApJS, 170, 175

\bibitem[\protect\citeauthoryear{Comastri et al.}{2002}]{cometal02} 	
Comastri, A., Mignoli, M., Ciliegi, P., et al. 2002, ApJ, 571, 771

\bibitem[\protect\citeauthoryear{Dickey \& Lockman}{1990}]{dicloc90} Dickey,
J.M., \& Lockman, F.J. 1990, ARA\&A, 28, 215 

\bibitem[\protect\citeauthoryear{Ezuka \& Ishida}{1999}]{ezuish99}
Ezuka, H., \& Ishida, M. 1999, ApJS, 120, 277 

\bibitem[\protect\citeauthoryear{Gilli et al.}{2007}]{giletal07}
Gilli, R., Comastri, A., \& Hasinger 2007, A\&A, 463, 79

\bibitem[\protect\citeauthoryear{Gonzalez et al.}{2005}]{gonetal05}
Gonzalez, A.H., Zabludoff, A.I., \& Zaritsky, D. 2005, ApJ, 618, 195

\bibitem[\protect\citeauthoryear{Grimm et al.}{2002}]{grietal02}
Grimm, H.-J., Gilfanov, M., \& Sunyaev, R. 2002, A\&A, 391, 923

\bibitem[\protect\citeauthoryear{Heckman et al.}{2005}]{hecetal05}
Heckman, T.M., Ptak, A., Hornschemeier, A., \& Kauffmann, G. 2005,
ApJ, 634, 161

\bibitem[\protect\citeauthoryear{Kalberia et al.}{2005}]{kaletal05}
Kalberla, P.M.W., Burton, W.B., Hartmann, et al. 2005, A\&A, 440,
775

\bibitem[\protect\citeauthoryear{Krivonos et al.}{2007}]{krietal07}
Krivonos, R., Revnivtsev, M., Lutovinov, A., Sazonov, S., Churazov,
E., \& Sunyaev, R. 2007, A\&A, 475, 775

\bibitem[\protect\citeauthoryear{Malizia et al.}{2007}]{maletal07}
Malizia, A., Landi, R., Bassani, L., et al. 2007, ApJ, 668, 81

\bibitem[\protect\citeauthoryear{Matt et al.}{2000}]{matetal00}
Matt, G., Fabian, A.C., Guainazzi, M., Iwasawa, K., Bassani, L., \&
Malaguti, G. 2000, MNRAS, 318, 173

\bibitem[\protect\citeauthoryear{Masetti et al.}{2006a}]{masetal06a}
Masetti, N., Bassani, L., Bazzano, A., Bird, A.J., \& Dean, A.J. 2006a,
A\&A, 455, 11

\bibitem[\protect\citeauthoryear{Masetti et al.}{2006b}]{masetal06b}
Masetti, N., Morelli, L., Palazzi, E., et al. 2006b, A\&A, 459, 21

\bibitem[\protect\citeauthoryear{Masetti et al.}{2008}]{masetal08}
Masetti, N., Mason, E., Morelli, L., et al. 2008, A\&A, 482, 113
	
\bibitem[\protect\citeauthoryear{Negueruela}{2004}]{negueruela04} 
Negueruela, I. 2004, in The Many Scales of the Universe -- JENAM 2004
Astrophysics Reviews (Kluwer Academic Publishers), ed. J. C. del Toro
Iniesta et al. [arXiv:astro-ph/0411759] 
	
\bibitem[\protect\citeauthoryear{Norberg et al.}{2002}]{noretal02} 
Norberg, P., Cole, S., Baugh, C.M., et al. 2002, MNRAS, 336, 907

\bibitem[\protect\citeauthoryear{Patterson}{1994}]{patterson94}
Patterson, J. 1994, PASP, 106, 209

\bibitem[\protect\citeauthoryear{Revnivtsev et al.}{2004}]{revetal04}
Revnivtsev, M., Sazonov, S., Jahoda, K., \& Gilfanov, M. 2004,
A\&A, 418, 927

\bibitem[\protect\citeauthoryear{Rodriguez et al.}{2008}]{rodetal08}
Rodriguez, J., Tomsick, J.A., \& Chaty, S. 2008, A\&A, 482, 731

\bibitem[\protect\citeauthoryear{Sazonov et al.}{2005}]{sazetal05}
Sazonov, S., Churazov, E., Revnivtsev, M., Vikhlinin, A., \& Sunyaev,
R. 2005, A\&A, 444, L37

\bibitem[\protect\citeauthoryear{Sazonov et al.}{2007}]{sazetal07}
Sazonov, S., Revnivtsev, M., Krivonos, R., Churazov, E., \& Sunyaev,
R. 2007, A\&A, 462, 57

\bibitem[\protect\citeauthoryear{Sazonov et al.}{2008}]{sazetal08}
Sazonov, S., Krivonos, R., Revnivtsev, M., Churazov, E., \& Sunyaev,
R. 2008, A\&A, 482, 517
	
\bibitem[\protect\citeauthoryear{Schlegel et al.}{1998}]{schetal98}
Schlegel, D.J., Finkbeiner, D.P., \& Davis, M. 1998, ApJ, 500, 525

\bibitem[\protect\citeauthoryear{Sugizaki et al.}{2001}]{sugetal01}
Sugizaki, M., Mitsuda, K., Kaneda, H., et al. 2001, ApJS, 134, 77

\bibitem[\protect\citeauthoryear{Suleimanov et al.}{2005}]{suletal05}
Suleimanov, V., Revnivtsev, M., \& Ritter, H. 2005, A\&A, 435, 191

\bibitem[\protect\citeauthoryear{Tueller et al.}{2008}]{tueetal08}
Tueller, J., Mushotzky, R.F., Barthelmy, S., et al. 2008, ApJ, 681 (in
press); arXiv0711.4130

\bibitem[\protect\citeauthoryear{Ubertini et al.}{2003}]{ubeetal03}
Ubertini, P., Lebrun, F., Di Cocco, G., et al. 2003, A\&A, 411, L131 

\bibitem[\protect\citeauthoryear{Ueda et al.}{2007}]{uedetal07} Ueda,
Y., Eguchi, S., Terashima, Y., et al. 2007, ApJ, 664, L79

\bibitem[\protect\citeauthoryear{Vikhlinin et al.}{1998}]{viketal98}
Vikhlinin, A., McNamara, B.R., Forman, W., Jones, C., Quintana, H., \&
Hornstrup, A. 1998, ApJ, 502, 558

\bibitem[\protect\citeauthoryear{Vikhlinin et al.}{2005}]{viketal05} 	
Vikhlinin, A., Markevitch, M., Murray, S.S., Jones, C., Forman, W., \&
Van Speybroeck, L. 2005, ApJ, 628, 655

\bibitem[\protect\citeauthoryear{Voges et al.}{1999}]{vogetal99} 
Voges, W., Aschenbach, B., Boller, Th., et al. 1999, A\&A, 349, 389

\bibitem[\protect\citeauthoryear{Winkler et al.}{2003}]{winetal03}
Winkler, C., Courvoisier, T.J.-L., Di Cocco, G., et al. 2003, A\&A,
411, L1 
	
\end{thebibliography}
\end{document}